# Domain-adversarial training of multi-speaker TTS

*Sunghee Jung, Hoirin Kim*

School of Electrical Engineering, KAIST, Daejeon, South Korea
`{sh.ee, hoirkim}@kaist.ac.kr`

## Abstract

Multi-speaker TTS has to learn both linguistic embedding and text embedding to generate speech of desired linguistic content in desired voice. However, it is unclear which characteristic of speech results from speaker and which part from linguistic content. In this paper, text embedding is forced to unlearn speaker dependent characteristic using gradient reversal layer to auxiliary speaker classifier that we introduce. We train a speaker classifier using angular margin softmax loss. In subjective evaluation, it is shown that the adversarial training of text embedding for unilingual multi-speaker TTS results in 39.9% improvement on similarity MOS and 40.1% improvement on naturalness MOS.

## 1. Introduction

There has been a lot of studies on text-to-speech(TTS) synthesis using deep neural networks[1][2]. Lately, there are attempts to train expressive TTS to demonstrate personality. Deep voice 2 and deep voice 3 are examples of multi-speaker TTS[3][4]. Unlike single speaker TTS, multi-speaker TTS has to disentangle speaker-dependent characteristics and linguistic content. Otherwise, the linguistic content is smoothed out and the clarity of the synthesized speech is lost due to different pronunciation styles of training speakers. There are studies on disentangling components of speech based on variational autoencoders and domain-adversarial neural networks[5][6][7]. Domain adversarial training is suggested for the purpose of unlearning bias that exists in the training data which is undesirable[8]. [5] and [6] uses domain-adversarial approaches introduced in [8] for TTS. [5] uses domain adversarial training to unlearn speaker-dependent characteristic when extracting text embedding for improving the speaker similarity of cross-lingual TTS, while [6] uses domain adversarial training for unlearning the existence of noise when extracting noise-robust speaker embeddings. [5] argues that adversarial learning resulted in higher similarity and lower naturalness when they tried to synthesized a speech in a language other than the language of the training speaker. It is intuitive that disentangling speaker and text becomes more crucial in such task.

When designing a speaker classifier for domain-adversarial learning, there are many implementational choices since speaker classifier is one of the most widely studied field in speech science. They can be categorized by how the speaker embeddings are extracted and what type of loss is used. I-vector used to be very popular for the representation of speaker, but now DNN-based embedding vectors are more widely studied. For loss, roughly speaking, variations of PLDA, triplet loss and softmax loss are widely studied[9][10][11][12][13]. For triplet loss, it has implementational drawback because minibatch has to be sampled across GPUs[11]. That is because it gets extremely hard to select negative samples in a GPU as the epoch increases. Unlike triplet loss, softmax-based losses have its advantage in implementation since there are no negative sampling issues. Softmax has different issue that it does not force inter-class distance to be large. However, angular softmax margin loss and additive angular softmax margin loss has been suggested to overcome this issue[12][13] [14][15]. In this study, we choose angular softmax margin loss for the speaker classifier. Detailed explanations on domain-adversarial training and angular softmax margin loss are given in next chapter.

## 2. Baseline studies

### 2.1. Domain adversarial neural network

Domain adversarial training of neural network is introduced in [8]. It is based on the idea that in order not to be affected by undesirable bias present in the training dataset, one can learn to classify such bias feature and implement negative gradient flow from the classifier to the feature extractor layer. It can be implemented by adopting gradient reversal layer between the feature extractor and the classifier. The gradient reversal layer can be explained as eq. (1).

$$GRL(x) = \begin{cases} x, & if\ forward\ path \\ -\lambda \times x, & if\ backward\ path \end{cases} \quad (1)$$

Here, $\lambda$ is a scaling factor for the gradient. According to [5], gradient reversal layer can cause unstable learning when the gradient from classifier is too large for an outlier data point. To prevent such case, authors of [5] uses gradient clipping techniques and limits the maximum value of gradient from the classifier network.

### 2.2. Angular softmax margin

Variations of softmax loss has been introduced in face recognition task first such as SphereFace and ArcFace[15][14]. That type of losses are of growing popularity in speaker recognition tasks as well due to its low equal error rate(EER) and simple implementation[12][13]. The angular softmax margin loss used in this paper is as in eq. (2).

$$L_{AMS} = -\frac{1}{n}\sum_{i=1}^{n} \log \frac{e^{s(\cos\theta_{y_i}-m)}}{e^{s(\cos\theta_{y_i}-m)} + \sum_{j=1, j\neq y_i}^{C} e^{s(\cos\theta_j)}} \quad (2)$$

If weight vector and feature vector are normalized to have size 1, eq.(2) can be written as eq.(3).

$$L_{AMS} = -\frac{1}{n}\sum_{i=1}^{n} \log \frac{e^{s(W_{y_i}^T f_i - m)}}{e^{s(W_{y_i}^T f_i - m)} + \sum_{j=1, j\neq y_i}^{C} e^{s(W_j^T f_i)}} \quad (3)$$

The authors of [11] and [12] argues that feature normalization is key feature to speaker recognition task since in this way, only the angular distance is used to measure the distance between weight vector and the feature.

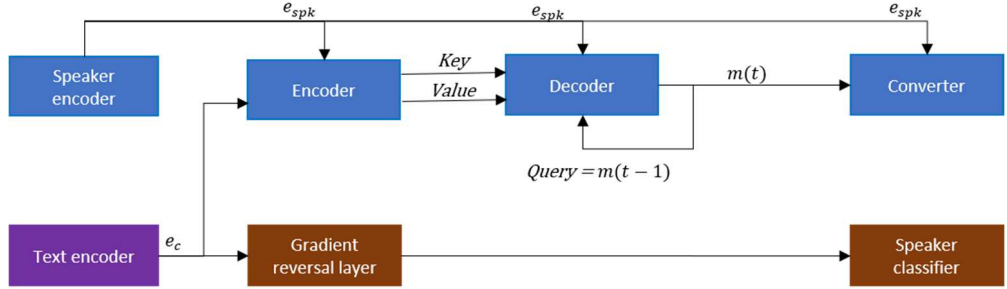

Figure 1. Architecture of proposed TTS system using adversarial learning

## 3. Overview of proposed idea

Figure 1 is the overview of the proposed multi-speaker TTS system with adversarial learning. $m(t)$ represents mel spectrogram at time t. $e_{spk}$ and $e_c$ represent speaker embedding and text embedding, respectively. The term text embedding and linguistic embedding are used interchangeably. TTS part is trained with L1 generation loss with ground truth mel spectrogram. Angular softmax margin loss in equation (3) is used to train speaker classifier. The entire model is trained jointly as in [5]. The gradient reversal layer has no effect for forward path, while it flows negative gradient in the backward path.

The TTS system is based on Deep voice 3 [4]. This system is characterized by the fact that it has speaker embedding injected in every convolutional layer of encoder, decoder and converter as condition, unlike many other researches on multi-speaker TTS where they simply append speaker embedding at the end of every text embedding at every time step [16][17][18][19]. This is also characteristic in that it has no RNN structure. CNN-based self-attention with positional encoding implemented in decoder replaces RNN and enables parallel computation over timesteps. This has advantage in training speed compared to RNN-based methods. On the other hand, auto-regressive structure is still used for the inference step because previous predicted mel spectrogram is used as query to the decoder as demonstrated in figure 1. In the training step, teacher-forcing releases this issue and enables parallel computation over multiple timesteps using convolutional neural network. More detailed explanation on baseline architecture can be found from [4].

## 4. Experiments and results

### 4.1. Dataset and feature extraction

English multi-speaker dataset LibriTTS [20] is used for training. Total 460 hours of clean training set is sued and the number of training speakers were 115. The dataset was trimmed based on the energy level of 60 dB at the beginning and the end of the sentences. The input of text is in grapheme as is originally given in the dataset. The audios were down-sampled to 22,050 Hz sampling rate and STFT is performed with 1024 frame window size and 256 frame hopping size. Then mel spectrograms were extracted using 80 mel bins.

### 4.2. Detailed model architecture

The architecture of baseline TTS is the same as in the deep voice 3. Speaker embedding dimension for LibriTTS is chosen as 256 dimensions while it was 512 in [4]. The architecture of the speaker classifier is as in [5] except that instead of softmax loss, we used angular margin softmax. The classifier has one hidden layer with 256 units. The scaling factor for the angular margin softmax loss is 40 and the angular margin is 0.6 as in [11]. The classifier tries to identify among 1151 speakers given the text embedding of 256 dimensions. λ of eq. (1) for negative scaling of gradient from speaker classifier to text encoder is 1.

### 4.3. Optimization

Adam optimizer is used with beta1 as 0.5, beta 2 as 0.9 and epsilon as e^-6 for hyper parameters. Initial learning rate is 0.0005 and learning rate is scheduled to decay with Noam learning rate decay[21]. Gradients are clipped if they were bigger than 0.1. Baseline paper [5] addresses that without gradient clipping, the training of text encoder can be unstable due to outliers in the speaker classifier.

### 4.4. Experiments

Naturalness MOS and similarity MOS was measured for the case with/without domain adversarial learning. For waveform generation, WORLD vocoder was used. The MOS test result is given in Table 1. In the subjective MOS evaluation, 6 participants were asked to assess 4 baseline sentences, 4 proposed sentences and one ground truth sentence from human target speaker. For measuring similarity MOS, participants were asked to compare the similarity of each synthesized sentence and target speaker ground truth speech. For naturalness MOS, participants were asked to evaluate how understandable and clear the contents were.

Table 1. Similarity and naturalness MOS. 'Baseline' is without domain adversarial training and 'proposed' is with domain adversarial training.

|  | Similarity | Naturalness |
|---|---|---|
| Baseline | 2.38 | 2.17 |
| Proposed | 3.33 | 3.04 |

### 4.5. Result analysis

In [5], domain-adversarial training helped disentangling linguistic embedding and speaker embedding under multi-lingual, multi-speaker TTS setting where only one speaker is available for each language and the goal was to synthesize different language than the original language of each voice (cross-lingual synthesis). Under such circumstances, it is very likely that the encoder would consider the speaker characteristic of speech as tightly coupled to the linguistic information. By training text encoder adversarial to speaker classifier, they were able to synthesize different language with the voice of speaker who originally spoke other language and demonstrated better similarity yet lower naturalness. It seemed the articulation accuracy of text encoder is deteriorated by the regularization of domain adversarial term from speaker classifier under cross-lingual setting. In such setting, each language has separate phoneme set and it is not shared across languages. Phoneme sets are simply concatenated to form united phoneme set. Notice that under that circumstance, each phone representation is only pronounced by one speaker.

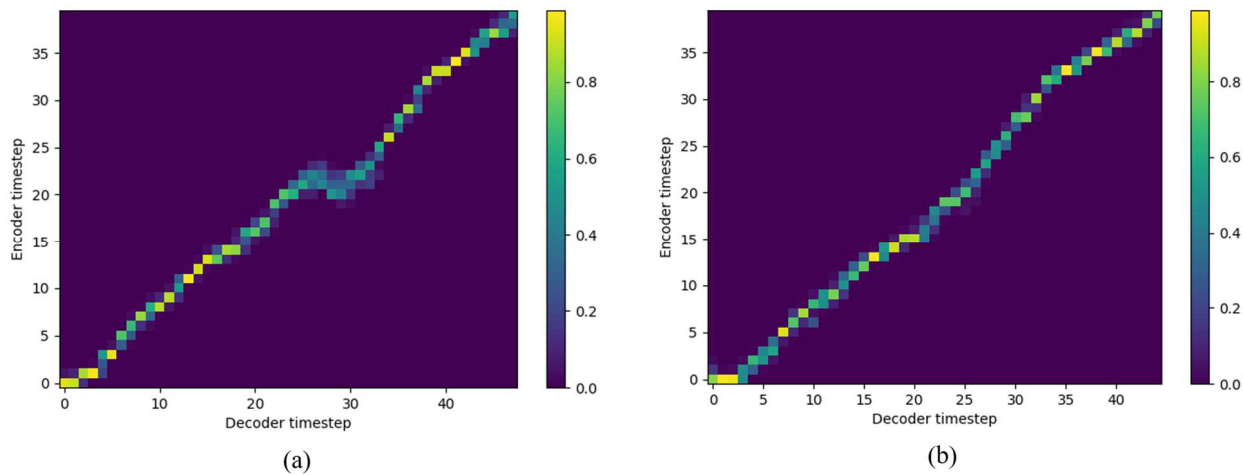

(a)                                                                (b)

Figure 2. Alignments of baseline and proposed system for *"Have a lovely evening with your friend."* (a) baseline system where text embedding is learned without domain-adversarial learning from speaker classifier. (b) proposed system where text embedding is learned

In this paper, however, domain-adversarial training also improved the naturalness not only the speaker similarity. The main reason why same technique gives such contrasting result is that, here, multiple people pronounces each phone representation in this paper. Under unilingual multi-speaker setting, phone representations are shared across all speakers while each of them pronounces it in a different way depending on their regional background. As a result, the linguistic embedding of each phoneme is smoothed out, not knowing whether to capture the difference in pronouncing style of same phone with linguistic embedding or speaker embedding. With the help of adversarial training to speaker classifier, such differences in pronouncing style are forced to be captured by speaker embedding. That is because text embedding is trained in a way that it should be oblivious of speakers by eq. (1).

However, for the case of multi-lingual TTS where one speaker speaks one language, the case where linguistic embedding is smoothed out by multiple pronouncing styles of phones by multiple training speakers does not happen. Thus, disentangling text embedding and speaker embedding does not benefit text embedding nor the naturalness MOS under cross-lingual setting.

It is observed in unofficial listening experience that improved quality of linguistic embedding contributes to better attention alignment and further increases the robustness of synthesized speech to tricky sentences. Figure 2. (a) shows alignment failure observed in baseline system while figure2. (b) shows successful alignment. Figure 2 (a) and (b) are both saying "Have a lovely evening with your friend." and the alignment error in (a) occurred at 'w' of 'with'. It is known that semi-vowels such as 'w' and 'y' varies in sound by noticeable amount depending on person and the context. It seems that learning linguistic embedding independent of speaker made it possible to learn robust representation of 'w' and resulted in robust alignment. It can be observed in figure2. (a) that even though alignments are unstable at local points, the TTS manages to recover its alignment in a few timestep., That is because monotonous alignment is forced so that when aligning encoder time steps and decoder timesteps, the decoder only attends 1 step backward and 3 steps forward.

## 5. Conclusions

In a subjective evaluation, speaker similarity was improved by 39.9% and naturalness was improved by 40.1% by training text embedding adversarial to speaker classifier. Also, it was observed qualitatively that with improved text embedding, it was able to obtain better alignment of encoder time steps and decoder time steps, generating robust synthesized speech.